\documentstyle[pre,aps]{revtex}
\begin{document}
\title{Distinguishing Fractional and White Noise in One and Two Dimensions}
\author{Alex Hansen\footnote{Permanent Address: Department of Physics,
NTNU, N--7491 Trondheim, Norway}}
\address{NORDITA and Niels Bohr Institute, Blegdamsvej 17, DK--2100
Copenhagen {\O}, Denmark}
\author{Jean Schmittbuhl}
\address{Departement de G{\'e}ologie, UMR CNRS 8538,
Ecole Normale Sup{\'e}rieure,
24, rue Lhomond, F--75231 Paris C{\'e}dex 05, France}
\author{G.\ George Batrouni}
\address{Institut Non-Lin\'eaire de Nice, UMR CNRS 6618, 
Universit{\'e} de Nice-Sophia Antipolis, 1361 Route des Lucioles,
F--06560 Valbonne, France}
\date{\today}
\maketitle
\begin{abstract} 
We discuss the link between uncorrelated noise and Hurst exponent for one
and two-dimensional interfaces. We show that long range correlations
cannot be observed using one-dimensional cuts through two-dimensional
self-affine surfaces whose height distributions are characterized by a
Hurst exponent, $H$, lower than $-1/2$. In this domain, fractional and
white noise are not distinguishable.  A method analysing the correlations in 
two dimensions is necessary. For $H>-1/2$, a cross-over regime
leads to an systematic over-estimate of the Hurst exponent. 
\vspace{0.3cm}

\noindent
PACS number(s): 47.55.Mh, 05.40.+j, 47.53.+n
\end{abstract} 
\vskip0.5cm

Self-affine surfaces are abundant in Nature.  They are the bread and butter
of quantitative characterization of growth phenomena such as fracture
surfaces \cite{b97}, interface growth and roughening phenomena \cite{bs97}. 

A self-affine surface $h(x,y)$ is defined by its behavior under the
scale transformation \cite{f88}
\begin{equation}
\label{eq:yawn}
\cases{x\to\lambda x\;,\cr
       y\to\lambda y\;,\cr
       h\to\lambda^H h\;,\cr}
\end{equation}
where $H$ is the Hurst exponent.  

Most commonly, the Hurst exponent is in the interval $0\le H\le 1$.
For instance, fracture surfaces exhibit a Hurst exponent close to 0.8
\cite{b97}.  Sea floor topography is self affine with a Hurst exponent close
to 0.5 \cite{t97}.
When $H>1$, the surface is no longer asymptotically flat.  When $H<0$,
the roughness distribution of the surface is refered to as {\it
fractional noise.\/} Fractional noise is typically encountered in
Nature in quantities that depend on the local slope of the topography:
mechanical stresses, light scattering and fluid flow \cite{vr97,svr99}.
For instance, the stress field on the interface
between two rough elastic blocks forced into complete contact is a
fractional noise with Hurst exponent $H_\sigma$ being related to the
Hurst exponent of the rough surface, $H$, as $H_\sigma = H - 1$
\cite{hsbo00}.                                                        

In this letter we show that, for values of $H$ in the range
$[-1,-1/2]$, self affinity takes on very different character in one
and two dimensions.  If this difference is ignored, one may obtain
wrong results when analyzing experimental data, no matter what method one
uses for estimating $H$.  Numerous tools exist for measuring Hurst exponents 
in the range $0<H<1$. Few of these methods have been tested systematically
in the range $H<0$ \cite{svr95}. 

The power spectrum of a self-affine trace $h(x)$, characterized by 
a Hurst exponent $H$, is given in one dimension by
\begin{equation}
\label{eq:1dpower}
P(k)\sim \frac{1}{k^{1+2H}}\qquad \mbox{in one dimension}\;,
\end{equation}
while the power spectrum of a two-dimensional self-affine surface
$h(x,y)$, characterized by the same Hurst exponent is
\begin{equation}
\label{eq:2dpower}
P(k)\sim \frac{1}{k^{2+2H}}\qquad \mbox{in two dimensions}\;.
\end{equation}

White, i.e., uncorrelated noise has a constant power spectrum
both in one and two dimensions.  Consequently, the value of the Hurst
exponent, $H$, which describes white noise in one dimension is
obtained from Eq.\ (\ref{eq:1dpower})
\begin{equation}
\label{eq:1dwhite}
H_{wn}=-\frac{1}{2}\qquad \;,
\end{equation}
while from Eq.\ (\ref{eq:2dpower}), we find for the two dimensional case
\begin{equation}
\label{eq:2dwhite}
H_{wn}=-1\qquad \;.
\end{equation}
This result is unexpected: One would have expected the value of the
Hurst exponent corresponding to white noise to be independent of
dimension.

This result is even more paradoxical when we analyze {\it cuts\/}
through a two-dimensional self-affine surface. Suppose one is given a
two-dimensional surface with Hurst exponent $H=-1/2$ and is asked to
determine $H$.  Analyzing the {\it two-dimensional\/} power spectrum of
this surface will lead to $P(k)\sim 1/k$ --- a 1/f spectrum, while
analysing the power spectrum of {\it one-dimensional cuts\/} through the
surface yields white noise.  We illustrate this point in Figs.\
\ref{fig1} and \ref{fig2} where we show one-dimensional cuts through
two-dimensional surfaces with $H=-1/2$ and $-1$ respectively.  The
synthetic surfaces were generated using a Fourier technique \cite{t97,s98}.

Yet a third problem is seen when analyzing a two-dimensional
self-affine surface with Hurst exponent in the range $-1\le H \le
-1/2$.  Analysing the correlations in the surface using the
two-dimensional power spectrum yields the correct value $-1\le H \le
-1/2$.  However, analysing one-dimensional cuts through the two
dimensional surface using the one-dimensional power spectrum method or
the average wavelet coefficient (AWC) method \cite{mrs97,shn98} yields
the {\it constant} value $H=-1/2$.  This is illustrated in Fig.\
\ref{fig3}.  On the other hand, analysing one-dimensional traces
generated with the Hurst exponent in the range $-1\le H \le -1/2$,
yields the input value of $H$.  This is illustrated in Fig.\
\ref{fig4}.  

This unexpected situation was recently encountered in the analysis of
the stress field of elastic self-affine surfaces in full contact
\cite{hsbo00}.  As mentioned above, if the elastic surfaces are
characterized by a Hurst exponent $H$, the corresponding stress field
has a Hurst exponent $H_\sigma=H-1$.  However, when analyzing the
stress using one-dimensional cuts, $H_\sigma$ was always saturating at
the value $-1/2$ as $H$ was lowered to values below $1/2$.

In order to understand what lies behind this unexpected behavior, we
need a model self-affine surface that is accessible to analytical
calculations. The model we choose is based on the Fourier method to
generate self-affine surfaces.

We discretize the surface, assuming it to be $h(n_x,n_y)$, where $0\le n_x
\le N-1$ and $0 \le n_y \le N-1$ are the positions of the nodes on a 
two-dimensional square lattice.  The surface may be represented in Fourier
space as
\begin{equation}
\label{eq:fsurf}
h(n_x,n_y)=\sum_{k_x=0}^{N-1} \sum_{k_y=0}^{N-1} e^{(2\pi i/N)[k_x n_x
+ k_y n_y]}\ \frac{\eta(k_x,k_y)}{(k_x^2+k_y^2)^{(H+1)/2}}\;,
\end{equation}
where $\eta(n_x,n_y)$ is a white (Gaussian) noise defined by a zero
mean and a second moment satisfying
\begin{equation}
\label{eq:white2}
\langle \eta(k_x,k_y)\eta(k'_x,k'_y)\rangle=2D\delta_{k_x,k'_x}
\delta_{k_y,k'_y}\;.
\end{equation}
We see immediately from Eq.\ (\ref{eq:fsurf}) that for $H=-1$,
$h(n_x,n_y)$ is white noise as we are then Fourier transforming the
white noise $\eta(k_x,k_y)$ directly.

A one-dimensional self-affine trace, on the other hand, may be written
\begin{equation}
\label{eq:ftrace}
h(n_x)=\sum_{k_x=0}^{N-1} e^{(2\pi i/N)k_x n_x}
\frac{\eta(k_x)}{k_x^{H+1}}\;,
\end{equation}
where $\eta(k_x)$ again is white noise.

In order to study a one-dimensional cut through the two-dimensional
surface $h(n_x,n_y)$, we place the cut along the $x$-axis and Fourier
transform $h(n_x,n_y)$ in the $x$-direction only.  This gives us
\begin{equation}
\label{eq:fhalf}
\tilde h(k_x,n_y)=\sum_{k_y=0}^{N-1}\ e^{(2\pi i/N)k_y n_y}\
\frac{\eta(k_x,k_y)}{(k_x^2+k_y^2)^{(H+1)/2}}\;.
\end{equation}
>From this expression, we readily construct the power spectrum along the
cut $n_y=constant$,  
\begin{equation}
\label{eq:1dpow}
P_y\left(k_x\right)=|\tilde h(k_x,0)|^2+|\tilde h(N-k_x,0)|^2\;,
\end{equation}
where we for simplicity and without loss of generality, have set
$n_y=0$.  Using Eq.\ (\ref{eq:white2}), we find
\begin{equation}
\label{eq:1dpow2}
P_y(k_x)=\frac{2D}{N^2}\ \sum_{k=0}^{N-1}\
\left[\frac{1}{(k_x^2+k^2)^{1+H}}+\frac{1}{(k_x^2+(N-k)^2)^{1+H}}\right]\;.
\end{equation}
For large $N$, this equation may be simplified to
\begin{equation}
\label{eq:1dpow3}
P_y(k_x)=\frac{2D}{N^2}\ \frac{1}{k_x^{1+2H}}\ 
\int_0^{N/k_x}\ \frac{dz}{(1+z^2)^{1+H}}\;.
\end{equation}
For $H>-1/2$, the integral in this equation approaches a constant rapidly
as $N\to\infty$.  However, for $H\le-1/2$, it behaves as
$(k_x/N)^{1+2H}$ for large $N$.  Thus, we conclude that
\begin{equation}
\label{eq:2dsolut}
P_y(k_x)\sim\cases{
(1/k_x)^{1+2H} & for $H>-1/2$,\cr
constant       & for $H\le -1/2$,\cr}
\end{equation}
This is precisely the behavior we see in Fig.\ \ref{fig3}.  On the other hand,
the power spectrum we find for the one-dimensional surface, Eq.\ 
(\ref{eq:ftrace}) is simply the one of Eq.\ (\ref{eq:1dpower}) irrespective
of the Hurst exponent $H$.

One important lesson we draw from this problem and its resolution is that
the Hurst exponent does {\it not\/} fully describe the correlations of
self-affine surfaces: A two-dimensional surface with a given Hurst exponent
may have completely different correlations from a one-dimensional surface
provided the Hurst exponent is low enough.  

Another important, but related lesson, is that measuring the self-affine
properties of a surface by averaging over one-dimensional cuts --- which is
the standard experimental approach --- may lead to wrong results.  In fact,
it was {\it knowing\/} the correct scaling of the stress field studied in
Ref.\ \cite{hsbo00} and comparing this to the measured quantities that 
led to this work.  Two-dimensional surfaces should preferably be analyzed
using two-dimensional tools.

We thank Kim Sneppen for useful discussions. This work was partially
funded by the CNRS PICS contract $\#753$ and the Norwegian research
council, NFR.  We also thank NORDITA for its hospitality and further
support.


\begin{figure}
\caption{A one-dimensional cut through a two-dimensional self-affine 
surface with $H=-1/2$.}
\label{fig1}
\end{figure}
\begin{figure}
\caption{A one-dimensional cut through a two-dimensional self-affine 
surface with $H=-1$.}
\label{fig2}
\end{figure}
\begin{figure}
\caption{Measured Hurst exponent $H_{mes}$ {\it vs.\/} Hurst exponent $H$
for two-dimensional surfaces.  Circles are based on power spectra measurements
along one-dimensional cuts, stars are based on AWC analysis along 
one-dimensional cuts, and filled lozenges are based on two-dimensional
power spectra measurements.}
\label{fig3}
\end{figure}
\begin{figure}
\caption{Measured Hurst exponent $H_{mes}$ {\it vs.\/} Hurst exponent $H$
for one-dimensional traces.  Circles are based on power spectra measurements
in one dimension, and stars are based on AWC analysis in one dimension.}
\label{fig4}
\end{figure}
\end{document}